\documentclass{article}\usepackage{amssymb,amsmath,amsthm,stmaryrd,mathrsfs,bm,verbatim}

\def\ip{\,{\vbox{\hrule height0pt\hbox{%
   \vrule height1.3ex width0pt\hskip1.2ex
   \vrule width0.3pt}\hrule height0.3pt\normalsize
  }} \, }

\theoremstyle{definition}

\newfont{\Fr}{eufm10}
\newfont{\Sc}{eusm10}
\newfont{\Bb}{msbm10}

\usepackage{hyperref}
\begin{document}

\title{A path integral derivation of $\chi_y$-genus}
\author{Guowu Meng\\ \small{\it Department of Mathematics}\\ \small{\it Hong Kong University of Science and Technology}\\
\small{\it Clear Water Bay, Kowloon, Hong Kong}\\
\small{Email: mameng@ust.hk}}
\maketitle

\begin{abstract} The formula for the Hirzebruch $\chi_y$-genus of complex manifolds is a
consequence of the Hirzebruch-Riemann-Roch formula. The classical index formulae for Todd genus,
Euler number, and Signature correspond to the case when the complex variable $y=$ $0$, $-1$, and
$1$ respectively. Here we give a {\it direct} derivation of this nice formula based on
supersymmetric quantum mechanics.
\end{abstract}



\section{Introduction}

It has been known for about twenty years that the Atiyah-Singer formula\cite{AS63} for the index of Dirac
operator can be directly derived from supersymmetric (SUSY) quantum mechanics \cite{lA83},
\cite{FW83}. Indeed, the well-known index formulae for Euler number, signature, Todd genus and
$\hat A$-genus can each be derived directly from a simple supersymmetric quantum
mechanic model.

In this paper we would like to point out that the formula for the Hirzebruch $\chi_y$-genus on
K\"{a}hler manifolds can also be derived directly from a simple supersymmetric
quantum mechanic model.

Let $M$ be a complex $d$-dimensional K\"{a}hler manifold, $b_{i,j}$ the $(i,j)$-Hodge number of
$M$, then the Hirzebruch $\chi_y$-genus is defined to be
\begin{eqnarray}
\chi_y(M)=\sum_{i,j}(-1)^iy^jb_{i,j}.
\end{eqnarray}

Let $\Omega=[\Omega_{{\bar
k}l}]_{1\le k, \hskip 3pt l\le d}$ be the curvature $2$-form of the K\"{a}hler metric,
$-i\Omega_n$ ($1\le n\le d$) the eigenvalue of antihermitian matrix $\Omega$,
$y$ a complex variable, $\sinh_y(x):={e^x+ye^{-x}\over 2}$. Then, the formulae for the Hirzebruch
$\chi_y$-genus is
\begin{equation}\label{E:F}
\chi_y(M)=\int_M\prod_n{\Omega_n/2\pi\over
\sinh(\Omega_n/4\pi)}\sinh_y (\Omega_n/4\pi)\left |_{\mbox{top form}}.\right.
\end{equation}
(Note that the integrand is a symmetric polynomial in $\Omega_1$, ..., $\Omega_d$, hence it is a
polynomial in traces of products of $\Omega$, i.e., a polynomial in Chern classes. Note also that
since both sides of (\ref{E:F}) are polynomials in $y$, we just need to derive (\ref{E:F}) for
$y$ being a complex number with unit norm.)

\section{SUSY Quantum Mechanics}

To begin with, following \cite{witten} (but with dimension reduced from $1+1$ to $0+1$), we
recall the standard supersymmetric quantum mechanics model on a K\"{a}hler manifold $M$ with
K\"{a}hler metric $g$. It governs maps
$\Phi$ :
$[0,\beta]\rightarrow X$, with $[0,\beta]$ being a close interval. If we pick
local coordinates $t$ on $[0,\beta]$ and $\phi^I$ on
$M$, then $\Phi$ can be described locally via functions $\phi^I(t)$. The local
complex coordinates on $M$ are denoted by $\phi^i$, and the complex conjugate
are denoted by $\phi^{\bar i}={\overline {\phi^i}}$. Let
$T'M$ be the holomorphic tangent bundle of $M$ and $T''M$ the antiholomorphic
tangent bundle of
$M$. The fermi fields of the model are $\psi^i_+$, $\psi^i_-$,
sections of $TM$, and $\psi^{\bar i}_+$, $\psi^{\bar i}_-$ sections of
$T''M$. The Lagrangian is

\begin{equation}\label{E:action}
I=\int_0^\beta dt\left ( g_{{\bar i}j}\dot\phi^{\bar i}\dot \phi^j
+ ig_{{\bar i}j}\psi_-^{\bar i}\nabla_t\psi^j_- + ig_{{\bar i}j}\psi_+^{\bar
i}\nabla_t\psi^{ j}_+  + R_{i{\bar j}k{\bar l}}\psi^{ i}_+\psi^{\bar
j}_+\psi^k_-\psi^{\bar l}_-
\right)
\end{equation}

The canonical quantization of this model gives rise to the
anticommutation relations among fermi fields at a fixed time $t$, the nontrivial
part of which is
\begin{equation}
\{\psi^i_-,\psi^{\bar i}_-\}|_t=2=\{\psi^i_+,\psi^{\bar i}_+\}|_t.
\end{equation}
Take $\psi^i_+(t)$, $\psi^{\bar i}_-(t)$ as the creation operators and
$\psi^i_-(t)$,
$\psi^{\bar i}_+(t)$ as the annihilation operators, then the Fock space that these
fermion operators generate is $\wedge^{*,*} (T_{\Phi(t)}M\otimes {\bf
C})$. So the Hilbert space of the model (denoted by ${\cal H}$) is the space of
complex differential forms, with
$\psi^i_-$ acting like ${i}dz^i\ip$,
$\psi^{\bar i}_-$ acting like ${i}dz^{\bar i}\wedge$, and  $\psi^i_+$ acting like
${i}dz^i\wedge$, $\psi^{\bar i}_+$ acting like ${i}dz^{\bar i}\ip$. This space admits
a ${\bf Z}\times {\bf Z}$-grading, given by operators $F_+$ and $F_-$, which
counts the plus-fermion numbers (holomorphic degree) and the minus-fermion
numbers (antiholomorphic degree) respectively.

Let $H$ be the Hamiltonian of the model, $q$ a complex number modulus one,
$\bar q$ the complex conjugate of $q$. A standard argument involving
coherent states \cite{bS81} shows that
\begin{equation}\label{E:ID}
Tr_{{\cal H}}\left((-1)^{F_-}(-{ q})^{F_+}e^{-\beta H}\right)=\int
e^{-S_E(\phi,\psi_+,\psi_-)}[d\phi][d\psi_-][d\psi_+]
\end{equation}
where the path integral is over configurations where $\phi$ and
$\psi_-$ are periodic and
$\psi^i_+(\beta)={ q}\psi^i_+(0)$ and $\psi^{\bar i}_+(\beta)={\bar q}\psi^{\bar
i}_+(0)$, and
\begin{equation} S_E=\int_0^\beta dt\left (  g_{{\bar i}j}\dot\phi^{\bar i}\dot
\phi^j + g_{{\bar i}j}\psi_-^{\bar i}\nabla_t\psi^j_- + g_{{\bar i}j}\psi_+^{\bar
i}\nabla_t\psi^{ j}_+  - R_{i{\bar j}k{\bar l}}\psi^{ i}_+\psi^{\bar
j}_+\psi^k_-\psi^{\bar l}_-
\right)
\end{equation}
is the Euclidean version of $I$ in equation (\ref{E:action})
obtained from $-iI$ by doing Wick-rotation:
\begin{equation}
i\beta\rightarrow \beta, \hskip 20pt it\rightarrow t.
\end{equation}

The supersymmetries of $I$ in (\ref{E:action}) that survive in the aforementioned
configurations are:
$$
\begin{array}{rclcrl}
\delta \phi^i & = & i\alpha \psi^i_-, &\hskip 20pt
\delta \phi^{\bar i} & = & i{\bar \alpha} \psi^{\bar i}_-\\
\\
\delta \psi^i_- & = & -{\bar \alpha} \dot\phi^i,  &\hskip 20pt
\delta \psi^{\bar i}_- & = & - \alpha \dot\phi^{\bar i}\\
\\
\delta \psi^{ i}_+ & = & -i\alpha\psi^j_-\Gamma_{jk}^i\psi^{ k}_+,
 &\hskip 20pt
\delta \psi^{\bar i}_+ & = & -i{\bar
\alpha}\psi^{\bar j}_-\Gamma_{{\bar j}{\bar k}}^{\bar i}\psi^{\bar k}_+
\end{array}
$$
And the corresponding supersymmetry charges are ${\bar\partial}+{\bar
\partial}^\dag$ (corresponding to
$\alpha$ being real) and
$-i(\bar \partial-{\bar \partial}^\dag)$ (corresponding to $\alpha$ being
imaginary). The space of supersymmetric invariant states (denoted by ${\cal
H}_0$) is identified with the space of complex harmonic forms. Then a standard
argument in SUSY quantum mechanics shows that the left hand side of
(\ref{E:ID}) is independent of $\beta$ and can be simplified as
\begin{equation}
Tr_{{\cal H}_0}\left((-1)^{F_-}(-q)^{F_+}\right)
\end{equation}
which is obviously the left-hand side of (\ref{E:F}) with $y=-q$.

It remains to show that the path integral in (\ref{E:ID}) can be simplified as
the right-hand side of (\ref{E:F}), and this will be done in the following
section.

\section{Evaluation of the Path Integral}
Since the path integral is independent of $\beta$, it is convenient to take the
limit $\beta\rightarrow 0$, when many higher-order interaction terms drop out
from the action. As $\beta\rightarrow 0$ the penalty for time variation in the
bosonic fields becomes larger and larger, and the nonconstant boson modes are
confined to narrower and narrower rangers; they are then approximated better and
better by tangent space variables. This allows the considerable simplification of
describing their contributions to the path integral as integrals over flat
vector spaces. The two main problems in calculating the functional integral are
to identify the leading terms as $\beta\rightarrow 0$ and to determine the
proper integration measure.

We will first consider the integration measure for the nonzero modes. This can
be determined from the free Euclidean action.

Let $q=e^{i\delta}$. Expanding the fields in Fourier components
\begin{eqnarray}
\phi^i={x_0}^i+\sqrt \beta\sum_{n\neq 0}^\infty
a_n^ie^{2\pi nt/\beta}, \cr
\phi^{\bar i}={x_0}^{\bar i}+\sqrt
\beta\sum_{n\neq 0}^\infty  a_n^{\bar i}e^{-2\pi nt/\beta},\cr
\psi^i_-={\sqrt{i\over 2\pi\beta}}\psi^i_0+\sum_{n\neq 0}
\psi_n^ie^{i2\pi nt/\beta},\cr
 \psi^{\bar i}_-={\sqrt{i\over 2\pi\beta}}\psi^{\bar
i}_0+\sum_{n\neq 0}
\psi_n^{\bar i}e^{-i2\pi nt/\beta},\cr
\psi^i_+=e^{i(\delta/4+\delta t/\beta)}\eta_0^i+\sum_{n\neq 0}
\eta_n^ie^{i(2\pi n+\delta)t/\beta},\cr
\psi^{\bar i}_+=e^{i(\delta/4-\delta t/\beta)}\eta_0^{\bar i}+\sum_{n\neq 0}
\eta_n^{\bar i}e^{-i(2\pi n+\delta)t/\beta}
\end{eqnarray}

The factors of $\sqrt \beta$ and $\sqrt {i\over 2\pi\beta}$ in the mode
expansions have the effect of removing all $\beta$-dependence from the
integration measure and setting the zero-mode measure to one.

We will integrate out first all nonzero modes of all fields, then all fermionic
zero modes, all of which are tangent-space variables, and finally we integrate
over $x_0^{\bar i}$ and $x_0^i$. Since the path integral is invariant under
changes in coordinates, we may perform the tangent space integrations using
normal coordinates centered at the corresponding point in $M$. This means that
we may set $x^i_0=0$ when performing these integrations.

Choosing a geodesic normal coordinate system around point $x_0$, the action reads
\begin{equation}
S_E=S_0+S_1+O(\beta)
\end{equation}
where

\begin{equation}
S_0=\sum_n\left[(2\pi n)^2a_n^{\bar i}a_n^i+i2\pi n(\psi_n^{\bar
i}\psi_n^i+\eta_n^{\bar i}\eta_n^i)\right ]
\end{equation}

and
\begin{equation}
 S_1=\sum_{n\neq 0}\left[i \Omega_{{\bar k}l}(i na_n^{\bar
k}a_n^l)-{i\Omega_{{\bar k}l}\over 2\pi}\eta_n^{\bar
k}\eta_n^l+i\delta\eta_n^{\bar i}\eta_n^i\right]+e^{i\delta/2}(-{i\Omega_{{\bar
k}l}\over 2\pi}\eta_0^{\bar
k}\eta_0^l+i\delta\eta_0^{\bar i}\eta_0^i)
\end{equation}
with $\Omega_{{\bar k}l}=R_{{\bar i}j{\bar k}l}\psi^{\bar i}_0\psi_0^j$.

The integration measure is
$$
[dx][d\psi_-][d\psi_+]=\prod_idx_0^idx_0^{\bar
i}d\psi_0^id\psi_0^{\bar i}d\eta_0^id\eta_0^{\bar i}d\mu^i
$$
where
$$
d\mu^i=\prod_{n\neq 0}({1\over
2\pi i}da_n^ida_n^{\bar i}d\psi_n^{ i}d\psi_n^{\bar i}d\eta_n^{i}d\eta_n^{\bar i})
$$
and is determined by requiring
\begin{equation}
\int \prod_i d\mu^i e^{-S_0}=1.
\end{equation}

Let $\Omega=[\Omega_{{\bar k}l}]_{1\le k, \hskip 3pt l\le d}$ be the antihermitian curvature
matrix, $I$ the identity matrix, integrating
$e^{-S_E}$ over
$\prod_i d\eta_0^id\eta_0^{\bar i}d\mu^i$, we get (in the limit $\beta\rightarrow 0$)
\begin{equation}\label{E:det}
\det \left (e^{i\delta /2}({i\Omega\over 2\pi}-i\delta I)\right ) \prod_{n>0}{\det}^{-1}\left (
I+{({i\Omega\over 4\pi})^2\over (n\pi)^2}\right ){\det}\left ( I+{({i\Omega\over
4\pi}-i{\delta\over 2}I)^2\over (n\pi)^2}\right )
\end{equation}

To evaluate the above expression, we assume $\Omega$ can be put in the diagonal form:

\begin{equation}\label{E:omega}
\Omega=\left (\begin{matrix}
-i\Omega_1 & &  \cr
&\ddots & \cr
&  & -i\Omega_d\end{matrix}\right ),
\end{equation}
then we get (using formula $\sinh x=x\prod_{n>0}\left (1+({x\over n\pi})^2\right )$)
\begin{equation}
\prod_i{{\Omega_i/ 2\pi}\over \sinh ({\Omega_i/4\pi})}\sinh_y({\Omega_i/4\pi})
\end{equation}
where $y=-e^{i\delta}$ and $\sinh_y(x)={e^x+ye^{-x}\over 2}$. This has a power series expansion
which terminates because of the Fermi statistics of the zero modes of $\psi_-$. Integrating it
over
$\prod_i d\psi_0^{ i}d\psi_0^{\bar i}$ projects out the coefficient of
$\prod_i\psi_0^{ i}\psi_0^{\bar i}$ in the integrand, i.e., the top form, which is
proportional to the volume form on $M$. For the form (\ref{E:omega}) of $\Omega$, the path
integral reduces to

\begin{equation}
\int_M \prod_i{{\Omega_i/ 2\pi}\over \sinh
({\Omega_i/4\pi})}\sinh_y({\Omega_i/4\pi})\left |_{\mbox{top form}}\right .
\end{equation}

Consider more general forms for $\Omega$, we note that expression (\ref{E:det}) can be seen as a
sum of traces of products of curvature forms, i.e., a polynomial in Chern classes. The splitting
principle then states that to calculate this polynomial, we may specialize to the case evaluated
explicitly above, where the vector bundle in question splits into a sum of line bundles.


\begin{thebibliography}{99}

\bibitem{AS63} M. F. Atiyah and I. M. Singer, Bull. Am. Math. Soc. {\bf {69}} (1963), 422; M. F.
Atiyah and I. M. Singer, Ann. Math. {\bf {87}} (1968), 484; ibid., 546.

\bibitem{lA83} L. Alvarez-Gaume, {\em Supersymmetry and the Atiyah-Singer index theorem}, Comm.
Math. Phys. {\bf {90}} (1983), 161.

\bibitem{FW83} D. Friedan and P. Windey, {\em Supersymmetric derivation of the Atiyah-Singer
index theorem and the chiral anomaly}, Nucl. Phys. {\bf {B235}} (1984), 395.

\bibitem{bS81} B. Sakita, {\em Quantum Theory of Many-Variable Systems and Fields}, World
Scientific, 1985
\bibitem{witten} E.Witten, {\em Mirror manifolds and topological field theory}, Essays on mirror
manifolds, 120--158, Internat. Press, Hong Kong, 1992.

\end{thebibliography}
\end{document}